\documentstyle[preprint,prl,aps]{revtex}

\begin{document}

\preprint{\vbox{ \hbox{July 1999}
\hbox{IFP-774-UNC} \hbox{BI-TP 99/18}
\hbox{hep-th/9907051}}}

\title{\bf Conformality from Field-String Duality
on Abelian Orbifolds}
\author{\bf Paul H. Frampton $^{(a,b)}$}
\address{ $^{(a)}$ 
Fakult\"at f\"ur Physik, Universit\"at Bielefeld, D-33501 Bielefeld, Germany.}
\address{$^{(b)}$ 
Department of Physics and Astronomy, University of North Carolina, Chapel Hill, NC  27599-3255, USA.
\footnote{Permanent address. 
e-mail: {\tt frampton@physics.unc.edu}} }

\maketitle

\begin{abstract}
If the standard model is embedded in a conformal theory, 
what is the simplest possibility? We
analyse all abelian orbifolds for discrete symmetry $Z_p$
with $p \leq 7$, and find that the simplest such theory is 
indeed $SU(3)^7$.  Such a theory predicts the correct electroweak 
unification (sin$^2 \theta \simeq 0.231$). A color 
coupling $\alpha_C(M) \simeq 0.07$ suggests a conformal scale $M$ near to 10 TeV.
\end{abstract}

\bigskip
\bigskip
\bigskip
\bigskip
\bigskip

\newpage

Since, in the context of field-string duality, there has been a shift
regarding the relationship of gravity
to the standard model of strong and electroweak interactions we
shall begin by characterising how gravity fits in,
then to suggest more 
specifically how the standard model fits in to the string framework.

The descriptions of gravity and of the standard model are contained
in the string theory. In the string picture in ten spacetime dimensions,
or upon compactification to four dimensions, there is a massless spin-two
graviton but the standard model is not manifest in the way we
shall consider it. In the conformal field theory
extension of the standard model, gravity is strikingly absent. 
The field-string duality does not imply that the standard model
already contains gravity and, in fact, it does not.

The situation is {\it not} analogous to the Regge-pole/resonance
duality (despite a misleading earlier version of this introduction!).
That quite different duality led to the origin\cite{Veneziano}
of string theory and originated from the realization\cite{IM,ARVV,DHS,BO}
phenomenologically that 
adding Regge pole and resonance descriptions is double counting
and that the two descriptions are {\it dual} in that stronger sense.
The duality \cite{Maldacena} between the field and
string descriptions is not analogous because the CFT
description does not contain gravity. 
A first step to combining gravity
with the standard model would be adding the
corresponding lagrangians. 

In the field theory description\cite{F1,FS,FV,F2}
used in this article, one will simply ignore the 
massless spin-two graviton. Indeed
since we are using the field theory description
only below the conformal scale of $\sim 1$TeV 
( or, as suggested later in
this paper, 10TeV) and forgoing any requirement of grand unification,
the hierarchy between the weak scale and theory-generated 
scales like $M_{GUT}$ or $M_{PLANCK}$ is resolved. 
Moreover, seeking 
the graviton in the field theory description is
possibly resolvable by going to a higher dimension
and restricting the range of the higher dimension.
Here we are looking only at the strong and weak interactions at accessible
energies below, say, 10TeV.

Of course, if we ask questions in a different regime, for example about
scattering of particles with center-of-mass energy of the
order $M_{PLANCK}$ then the graviton
will become crucial\cite{Hooft,ACV} and a string, rather than a 
field, description will be the viable one.

It is important to distinguish between the holographic 
description of the five-dimensional gravity
in $(AdS)_5$ made by the four-dimensional CFT and the origin of the
four-dimensional graviton. The latter could be described holographically
only by a lower three-dimensional field theory which is not
relevant to the real world. Therefore the graviton
of our world can only arise by {\it compactification} of a higher
dimensional graviton.
Introduction of gravity must break conformal 
invariance and it is an interesting 
question (which I will not answer!) whether
this breaking is related to the mass and 
symmetry-breaking scales in the low-energy theory. That is
all I will say about gravity in the present paper; the
remainder is on the standard model and its embedding in a CFT.

An alternative to conformality, grand unification with supersymmetry, 
leads to an impressively accurate gauge coupling unification\cite{ADFFL}.
In particular it predicts an electroweak mixing angle
at the Z-pole, ${\tt sin}^2 \theta = 0.231$. This result 
may, however, be fortuitous, but rather than
abandon gauge coupling unification, we can rederive ${\tt sin}^2 \theta = 0.231$
in a different way by embedding the electroweak $SU(2) \times U(1)$ in
$SU(N) \times SU(N) \times SU(N)$ to
find ${\tt sin}^2 \theta = 3/13 \simeq 0.231$\cite{FV,F2}.
This will be a common feature of the models in this paper.

The conformal theories will be finite without
quadratic or logarithmic divergences. This requires appropriate 
equal number of fermions and bosons which can cancel in
loops
and which occur without the necessity of space-time supersymmetry.
As we shall see in one example, it is possible to combine
spacetime supersymmetry
with conformality but the latter is the driving principle and the former
is merely an option: additional fermions and scalars are predicted by
conformality in the TeV range\cite{FV,F2}, 
but in general these particles are different and distinguishable
from supersymmetric partners.
The boson-fermion cancellation is essential for the
cancellation of infinities, and will
play a central role in the calculation of the cosmological constant
(not discussed here). In the field picture, the
cosmological constant measures the vacuum energy density.

What is needed first for the conformal approach is a simple
model and that is the subject of this paper.

Here we shall focus on abelian orbifolds characterised by the discrete group
$Z_p$. Non-abelian orbifolds will be systematically analysed elsewhere.

The steps in building a model for the abelian case (parallel steps
hold for non-abelian orbifolds) are:

\begin{itemize}

\item{(1)} Choose the discrete group $\Gamma$. Here we are considering
only $\Gamma = Z_p$. We define $\alpha = {\rm exp}(2 \pi i/p)$.

\item{(2)} Choose the embedding of $\Gamma \subset SU(4)$ by
assigning ${\bf 4} = (\alpha^{A_1}, \alpha^{A_2}, \alpha^{A_3}, \alpha^{A_4})$
such that $\sum_{q=1}^{q=4} A_q = 0 ({\rm mod} p)$. To
break ${\cal N} = 4$ supersymmetry to
${\cal N} = 0$ ( or ${\cal N} = 1$) requires that
none (or one) of the $A_q$ is equal to zero (mod p).

\item{(3)} For chiral fermions one requires that ${\bf 4} \not\equiv {\bf 4^{*}}$
for the embedding of $\Gamma$ in $SU(4)$.

The chiral fermions are in the bifundamental representations of $SU(N)^p$
\begin{equation}
\sum_{i=1}^{i=p} \sum_{q=1}^{q=4} 
(N_i, \bar{N}_{i + A_q})
\label{fermions}
\end{equation}
If $A_q=0$ we interpret $(N_i, \bar{N}_i)$ as a singlet plus an adjoint
of $SU(N)_i$.

\item{(4)} The {\bf 6} of $SU(4)$ is real {\bf 6} = $(a_1, a_2, a_3, -a_1, -a_2, -a_3)$
with 
$a_1 = A_1 + A_2$,
$a_2 = A_2 + A_3$,
$a_3 = A_3 + A_1$
(recall that all components are defined modulo p).
The complex scalars are in the bifundamentals
\begin{equation}
\sum_{i=1}^{i=p} \sum_{j=1}^{j=3} 
(N_i, \bar{N}_{i \pm a_j})
\label{scalars}
\end{equation}
\noindent The condition in terms of $a_j$ for ${\cal N} = 0$ is
$\sum_{j=1}^{j=3} (\pm a_j) \not= 0 ({\rm mod}~~ p)$\cite{F1}.
\item{(5)} Choose the $N$ of $\bigotimes_i SU(Nd_i)$ (where the $d_i$ are the
dimensions of the representrations of $\Gamma$). For the abelian case
where $d_i \equiv 1$, it is natural to choose $N=3$ the largest
$SU(N)$ of the standard model (SM) gauge group. For a non-abelian $\Gamma$
with $d_i \not\equiv 1$ the choice $N=2$ would be indicated.

\item{(6)}  The $p$ quiver nodes are identified as color (C), weak isospin (W)
or a third $SU(3)$ (H). This specifies the embedding of the gauge group
$SU(3)_C \times SU(3)_W \times SU(3)_H \subset \bigotimes SU(N)^p$.

This quiver node identification is guided by (7), (8) and (9) below.

\item{(7)}  The quiver node identification is required to
give three chiral families under Eq.(\ref{fermions})
It is sufficient to make three of the $(C + A_q)$ to be W and the fourth H, given that there is only
one C quiver node, so that there are three $(3, \bar{3}, 1)$. Provided that 
$(\bar{3}, 3, 1)$ is avoided by the $(C - A_q)$ being H, the remainder
of the three family trinification will be automatic by chiral anomaly cancellation.
Actually, a sufficient condition for three families has been given; it is
necessary only that the difference between the number of $(3 + A_q)$ 
nodes and the number of $(3 - A_q)$ nodes
which are W is equal to three.

\item{(8)}  The complex scalars of Eq. (\ref{scalars}) must be sufficient for their
vacuum expectation values (VEVs) to spontaneously break 
$SU(3)^p \longrightarrow SU(3)_C \times SU(3)_W \times SU(3)_H
\longrightarrow SU(3)_C \times SU(2)_W \times U(1)_Y \longrightarrow SU(3)_C \times U(1)_Q$.

Note that, unlike grand unified theories (GUTs) with or without supersymmetry,
the Higgs scalars are here prescribed by the conformality condition.
This is more satisfactory because it implies that the Higgs sector cannot be chosen
arbitrarily, but it does make model building more interesting

\item{(9)} Gauge coupling unification should apply at least to the electroweak mixing
angle ${\rm sin}^2 \theta = g_Y^2 / (g_2^2 + g_Y^2) \simeq 0.231$. For trinification
$Y = 3^{-1/2} ( - \lambda_{8W} + 2\lambda_{8H})$ so that $(3/5)^{1/2} Y$ is
correctly normalized. If we make $g_Y^2 = (3/5)g_1^2$ and $g_2^2 = 2 g_1^2$
then ${\rm sin}^2 \theta = 3/13 \simeq 0.231$ with sufficient accuracy.

\end{itemize}

\bigskip
\bigskip

In the remainder of this paper we answer all these steps for the choice $\Gamma = Z_p$ 
for successive $p = 2, 3 ...$ up to $p = 7$, then add some concluding remarks.

\bigskip
\bigskip

\begin{itemize}

\item{{\bf p = 2}}

In this case $\alpha = -1$ and therefore one cannot
costruct any complex {\bf 4} of $SU(4)$ with
${\bf 4} \not\equiv {\bf 4^*}$. 
Chiral fermions are therefore
impossible.

\item{{\bf p = 3}}

The only possibilities are $A_q = (1, 1, 1, 0)$ or $A_q = (1, 1, -1, -1)$.
The latter is real and leads to no chiral fermions.
The former leaves ${\cal N} = 1$ supersymmetry and is a simple three-family
model\cite{KS} by the quiver node identification C - W - H. The scalars $a_j = (1, 1, 1)$
are sufficient to spontaneously break to the SM. Gauge coupling unification is, however,
missing since ${\rm sin}^2 \theta = 3/8$, in
bad disagreement with experiment.

\item{{\bf p = 4}}

The only complex ${\cal N} = 0$ choice is $A_q = (1, 1, 1, 1)$. But
then $a_j = (2, 2, 2)$ and any quiver node identification
such as C - W - H - H has 4 families and the scalars are insufficient to
break spontaneously the symmetry to the SM gauge group.

\item{{\bf p = 5}}

The two inequivalent complex choices are $A_q = (1, 1, 1, 2)$ and $A_q = (1, 3, 3, 3)$.
By drawing the quiver, however, and using the rules 
for three chiral families given in (7)
above, one finds that the node identification and the prescription of the scalars
as $a_j = (2, 2, 2)$ and $a_j = (1, 1, 1)$ respectively does not permit
spontaneous breaking to the standard model.

\item{{\bf p = 6}}

Here we can discuss three inequivalent complex possibilities as follows:

\noindent (6A) $A_q = (1, 1, 1, 3)$ which implies $a_j = (2, 2, 2)$.

\bigskip

\noindent Requiring three families means a node identification C - W - X - H - X - H
where X is either W or H. But whatever we choose for the X the scalar representations
are insufficient to break $SU(3)^6$ in the desired fashion down to the SM. This
illustrates the difficulty of model building when the scalars are not
in arbitrary representations.

\noindent (6B) $A_q = (1, 1, 2, 2)$ which implies $a_j = (2, 3, 3)$.

\bigskip

\noindent 
Here the family number can be only zero, two or four as can be seen by inspection
of the $A_q$ and the related quiver diagram. So (6B) is of no phenomenological interest.

\noindent (6C) $A_q = (1, 3, 4, 4)$ which implies $a_j = (1, 1, 4)$.

\bigskip

\noindent Requiring three families needs a quiver node identification which is of the form
{\it either}
C - W - H - H - W - H {\it or} C - H - H - W - W - H. The scalar representations
implied by $a_j = (1, 1, 4)$ are, however, easily seen to be insufficient to do the required
spontaneous symmetry breaking (S.S.B.) for both of these identifications. 

\item{{\bf p =7}}

Having been stymied mainly by the rigidity of the scalar representation
for all $p \leq 6$, for $p = 7$ there
are the first cases which work. Six inequivalent complex embeddings of $Z_7
\subset SU(4)$ require consideration.

\noindent (7A) $A_q = (1, 1, 1, 4) \Longrightarrow a_j = (2, 2, 2)$

\bigskip

For the required nodes C - W - X - H - H - X - H the
scalars are {\it insufficient} for S.S.B.

\bigskip

\noindent (7B) $A_q = (1, 1, 2, 3) \Longrightarrow a_j = (2, 3, 3)$

\bigskip

The node identification C - W - H - W - H - H - H leads
to a {\it successful} model.

\bigskip

\noindent (7C) $A_q = (1, 2, 2, 2) \Longrightarrow a_j = (3, 3, 3)$

\bigskip

Choosing C - H - W - X - X - H - H to derive three families, the
scalars {\it fail} in S.S.B.

\bigskip

\noindent (7D) $A_q = (1, 3, 5, 5) \Longrightarrow a_j = (1, 1, 3)$

\bigskip

The node choice C - W - H - H - H - W - H leads
to a {\it successful} model. This is Model A of \cite{F2}.

\bigskip

\noindent (7E) $A_q = (1, 4, 4, 5) \Longrightarrow a_j = (1, 2, 2)$

\bigskip

The nodes C - H - H - H - W - W - H are 
{\it successful}.

\bigskip

\noindent (7F) $A_q = (2, 4, 4, 4) \Longrightarrow a_j = (1, 1, 1)$

\bigskip

Scalars {\it insufficient} for S.S.B.

\end{itemize}

The three successful models (7B), (7D) and (7E) lead to
an $\alpha_3(M) \simeq 0.07$. Since $\alpha_3(1 {\rm TeV}) \geq 0.10$
this suggest a conformal scale $M \simeq 10$ TeV \cite{F2}.
The above models have less generators than an
$E(6)$ GUT and thus $SU(3)^7$ merits
further study. It is possible, and under investigation, that non-abelian
orbifolds will lead to a simpler model.

\bigskip

For such field theories it is important to establish the existence of a fixed
manifold with respect to the renormalization group. It 
could be a fixed line but more likely, in
the ${\cal N} = 0$ case, a fixed point\cite{CST}. It is known that in the
$N \longrightarrow \infty$ limit the theories become conformal, but 
although this 't Hooft limit\cite{tH} is where the field-string duality
is derived we know that finiteness survives to finite N in the
${\cal N} = 4$ case\cite{mandelstam} and this makes it plausible
that at least a conformal point occurs also for the ${\cal N} = 0$
theories with $N = 3$ derived above.

The conformal structure cannot by itself predict all the dimensionless ratios of
the standard model such as mass ratios and mixing angles 
because these receive contributions,
in general, from soft breaking of conformality. With a 
specific assumption about the pattern of conformal
symmetry breaking, however, more work should lead to 
definite predictions for such quantities.

\bigskip
\bigskip
\bigskip
\bigskip
\bigskip
\bigskip

The hospitality of Bielefeld University
is acknowledged while this work was done. The work was
supported in part by the US Department of Energy under 
Grant No. DE-FG02-97ER-41036.

\end{document}